\documentclass[prd,onecolumn,nofootinbib]{revtex4}
\usepackage{bm,amsmath,amssymb,graphicx}

\begin{document}

\title{Universe in a rotating black hole and preferred axis}
\author{Nikodem Pop{\l}awski}
\altaffiliation{NPoplawski@newhaven.edu}
\affiliation{Department of Mathematics and Physics, University of New Haven, 300 Boston Post Road, West Haven, CT 06516, USA}

\begin{abstract}
If our universe was born as a baby universe on the other side of the event horizon of a black hole existing in a parent universe, then the corresponding white hole at rest provides the absolute frame of reference in the universe.
In this frame, the cosmic microwave background radiation is isotropic on large scales.
If the parent black hole is rotating, then its axis of rotation becomes a preferred axis in the universe.
Accordingly, the absolute frame is non-inertial, although the non-inertial forces are small.
To decrease their energies, galaxies tend to align their axes of rotation with the preferred axis, resulting in clockwise-counterclockwise asymmetry.
The centrifugal force causes a large-scale bulk flow of galaxy clusters in directions perpendicular to the preferred axis.
The astronomical data seem to support these motions.
The angular velocity of the universe decreases as the universe expands, which is a consequence of the conservation of the angular momentum of the universe.
The centrifugal force in a rotating universe, which also decreases, may be the origin of dark energy, in accordance with recent DESI observations showing that dark energy becomes weaker with time.
\end{abstract}
\maketitle

\noindent
{\bf Absolute frame of reference}.\\
If there were only one body in the universe, for example Earth, then it would have no relative motion with respect to other bodies.
Yet, a Foucault's pendulum could determine Earth's rotation about its own axis \cite{LL1}.
But, with respect to what would Earth rotate if there were no other bodies?
Earth's state of motion would have no meaning in that case.

Newton's rotating bucket argument demonstrated that true rotational motion cannot be defined as the relative rotation of the body with respect to the immediately surrounding bodies \cite{Newton}.
More generally, true motion and rest cannot be defined relative to other bodies. Instead, they can be defined only by reference to absolute space.

Einstein's general theory of relativity, in which the motion of bodies is determined by the local geometry of spacetime \cite{LL2}, reduces absolute space and time to local geodesics that are sufficient to describe this geometry.
Absolute space becomes a field that is described by the metric tensor.
True motion and rest are defined by reference to the metric tensor that asymptotically (far away from physical bodies) tends to the form determined by the condition of constant curvature, which depends on whether the universe is flat, closed, or open.

According to Einstein, the metric tensor is determined locally by the distribution of matter.
What determines the asymptotic form of the metric tensor that took the role of absolute space?
Mach's principle states that the overall distribution of matter provides absolute space: local physical laws are determined by the large-scale structure of the universe \cite{Mach}.
Consequently, the motion of the distant stars determines the local inertial frame.
But, if there were no bodies other than Earth, there would be no distant matter that could determine the metric tensor.
The metric would be redundant because it would have no physical bodies to relate to.
Yet, the metric, taking the role of absolute space, must exist in order to explain the difference between two scenarios in which the plane of oscillation of a Foucault’s pendulum rotates (indicating Earth's rotation with respect to the metric field) or not.
Therefore, there must exist a body in the universe that determines the absolute space.

I propose that black hole cosmology (BHC) provides a natural explanation of the absolute space \cite{cosmo,ApJ,shear}.
If our universe was born as a closed, baby universe \cite{BHG} formed on the other side of the event horizon of a parent black hole existing in a parent universe, then that black hole would be seen in the baby universe as a primordial white hole.
That white hole determines absolute space: the frame of reference in which the white hole is at rest is the absolute frame of reference (AFR).\footnote{
The white hole being at rest means that its comoving distance (fixed relative to growing space) from the observer does not change.
The physical distance grows because the universe is expanding.
}
This frame defines the absolute time, called the cosmic time, which appears in the Friedmann equations of cosmology.
It also defines absolute simultaneity and comoving distances.

Newton's first law of dynamics (the law of inertia) has two parts: 1) There exist inertial frames of reference, and 2) In an inertial frame, a free body (without forces acting on it) has a constant velocity: “a body at rest stays at rest and a body in motion stays in motion.”
In the Lagrangian formulation, an inertial frame of reference is a frame in which space is homogeneous and isotropic, and time is homogeneous \cite{LL1}.
If the parent black hole is not rotating, then the rest frame of reference of the corresponding primordial white hole is inertial.
The white hole guarantees that there is at least one inertial frame: the AFR, which explains the origin of inertia.
The principle of relativity of Galileo and Einstein allows then to construct, through the Lorentz transformations, an infinite number of other inertial frames that are mechanically equivalent to AFR and to each other \cite{LL2}.

I propose two conjectures.
Firstly, in the AFR, the total momentum and angular momentum of the matter in the universe are zero (they can be measured only within the observer's cosmological horizon).
These definitions are determinate because the universe formed by a black hole is closed (with the exception of the white hole that connects the universe to the parent universe through an Einstein--Rosen bridge).
This conjecture has the spirit of Mach's principle: all distant matter determines inertia \cite{Mach}.
Secondly, in the AFR, the cosmic microwave background (CMB) radiation is isotropic (on largest scales, without accounting for small temperature fluctuations that have led to large-structure formation) \cite{pref}.

Therefore, BHC completes Einstein's general theory of relativity by making it Machian.
The parent black hole, seen in our universe as a white hole, constitutes the Machian distant matter that determines absolute space and provides inertial frames of reference.

There are further implications of BHC.
If our universe was born on the other side of the event horizon of a black hole existing in a parent universe, then every black hole creates a new, baby universe.
These universes form a multiverse.
However, they are not parallel.
An object can exist at any moment of its timeline (measured in its proper time defined in its rest frame) only in one universe.

Since the motion of matter through an event horizon can only occur in one direction, that motion can define the past and future.
This existence of the arrow of time at the event horizon can be continuously extended to all other points in space.
It will also be extended to the cosmic time, defined as the time in the AFR, which is a coordinate that measures the expansion of the universe.
Accordingly, the cosmic arrow of time in the universe would be inherited from the parent universe in BHC \cite{cosmo}.

The second law of thermodynamics states that the total entropy of an isolated system can never decrease over time.
Once the total entropy of a system and its surroundings reaches a maximum, it would remain constant: the system is in thermodynamic equilibrium with the surroundings.
A black hole, which forms in the infinite future as measured in the AFR, is a cosmic example of such a system.
However, the AFR of the parent universe cannot be extended beyond the event horizon of a black hole because of the infinite redshift at the horizon \cite{LL2}.
On the other side of the event horizon, a baby universe has its own AFR and its own cosmic time.
In that growing universe, entropy can increase further \cite{ApJ}.
Therefore, the thermodynamic and cosmic arrows of time coincide.

The black hole information paradox does not exist.
The information goes from the parent universe to a baby universe formed on the other side of the black hole's event horizon.
Because the curvature of the closed universe is absolute, the gravitational force is geometrical and does not need a mediating particle; the graviton does not exist.

A physical law that turns black holes into Einstein--Rosen bridges to new, baby universes must avoid the black-hole singularity.
The simplest and most natural mechanism for preventing gravitational singularities is provided by spacetime torsion within the Einstein--Cartan theory of gravity \cite{Newton,EC}.
In this theory, torsion is coupled to the intrinsic angular momentum of fermionic matter, allowing for the spin-orbit interaction that follows from the Dirac equation.
Accordingly, torsion brings the consistency between relativistic quantum mechanics and curved spacetime.
At extremely high densities, torsion manifests itself as repulsive gravity, preventing the formation of a singularity and creating a big bounce that starts a new universe \cite{ApJ,avert}.
Quantum effects in a strong, time-varying gravitational field cause intense particle production, which overcomes the effects of shear that oppose torsion \cite{shear}.
This production also causes an exponential expansion of the new universe (inflation) after the big bounce, which lasts a finite time until torsion becomes weak \cite{ApJ}.
Torsion and particle production therefore provide the simplest and most natural explanation of inflation, without assuming the existence of hypothetical fields.
This origin of inflation is consistent with the Planck observations of the CMB \cite{obs}.

The inertia in the universe may therefore originate from the universe being formed by a black hole that naturally provides the AFR: the absolute frame of reference.
This formation is physically realized through initial gravitational attraction from curvature, which is later countered by gravitational repulsion from torsion.

In addition to eliminating gravitational singularities, torsion may also remove divergences in Feynman diagrams in quantum electrodynamics, resulting in finite values of bare (before renormalization) quantities such as the mass and electric charge of the electron \cite{reg}.
Torsion may be a physical mechanism, ensuring that all physics is finite.\\

\noindent
{\bf Rotating frame of reference in a rotating black hole}.\\
Because most stars rotate, most black holes are rotating black holes, described by the Kerr metric \cite{Kerr}.
They do not expand as a result of the expansion of the universe \cite{exp}; the universes created by them do expand, but this dynamics cannot be observed outside the event horizon.
A universe born from a rotating black hole should inherit its axis of rotation as a preferred axis \cite{cosmo}.
A complete description of a universe formed in a rotating black hole should combine the Friedmann--Lema\^{i}tre--Robertson--Walker (FLRW) metric, describing an expanding universe, with the G\"{o}del metric, describing a rotating universe \cite{rot}.
The preferred direction should introduce small corrections to the FLRW metric, containing the Kerr radius $a=M/mc$, where $M$ is the angular momentum of a rotating black hole and $m$ is its mass.

Several qualitative consequences of the rotation of the universe can be derived using non-relativistic mechanics \cite{LL1}.
In an inertial frame of reference $K_0$, the Lagrangian for a non-relativistic particle of mass $m$ with radius vector ${\bf r}_0$, moving with velocity ${\bf v}_0=d{\bf r}_0/dt$ in an external field represented by the potential energy $U$, is $L_0=\frac{1}{2}m{\bf v}^2_0-U({\bf r_0})$.
In a frame of reference $K$, which rotates relative to $K_0$ with angular velocity ${\bm\Omega}(t)$ and whose origin coincides with the origin of $K_0$, the velocity ${\bf v}=d{\bf r}/dt$ of the particle is given by ${\bf v}_0={\bf v}+{\bm\Omega}\times{\bf r}$ and ${\bf r}={\bf r}_0$.
Putting these relations into $L_0$ gives the Lagrangian of the particle in the rotating frame $K$:
\begin{equation}
L=\frac{1}{2}m{\bf v}^2+m{\bf v}\cdot({\bm\Omega}\times{\bf r})+\frac{1}{2}m({\bm\Omega}\times{\bf r})^2-U({\bf r}).
\end{equation}
Putting the partial derivatives:
\begin{equation}
\frac{\partial L}{\partial{\bf v}}=m{\bf v}+m{\bm\Omega}\times{\bf r},\quad \frac{\partial L}{\partial{\bf r}}=m{\bf v}\times{\bm\Omega}+m({\bm\Omega}\times{\bf r})\times{\bm\Omega}-\frac{\partial U}{\partial{\bf r}},
\end{equation}
into the Lagrange equation of motion $(d/dt)(\partial L/\partial{\bf v})-\partial L/\partial{\bf r}=0$ gives the Newton equation of motion in a rotating frame:
\begin{equation}
m\frac{d{\bf v}}{dt}=-\frac{\partial U}{\partial{\bf r}}-m{\bm\alpha}\times{\bf r}-2m{\bm\Omega}\times{\bf v}-m{\bm\Omega}\times({\bm\Omega}\times{\bf r}),
\end{equation}
where ${\bm\alpha}=d{\bm\Omega}/dt$ is the angular acceleration of $K$.
The effect of a rotational motion of a frame of reference with angular velocity ${\bm\Omega}$ is equivalent to the effect of three forces: the force $m{\bm\alpha}\times{\bf r}$ arising from non-uniformity of the rotation, the Coriolis force $-2m{\bm\Omega}\times{\bf v}$, and the centrifugal force $-m{\bm\Omega}\times({\bm\Omega}\times{\bf r})$ \cite{LL1}.
The centrifugal force acts in the direction opposite to ${\bf r}$ (away from the axis of rotation passing through the origin of $K$) and its magnitude is $m\Omega^2\rho$, where $\rho$ is the distance of the particle from the axis of rotation.

The momentum ${\bf p}=\partial L/\partial{\bf v}$ for the Lagrangian $L$ is
\begin{equation}
{\bf p}=m{\bf v}+m{\bm\Omega}\times{\bf r}=m{\bf v}_0={\bf p}_0.
\end{equation}
The momentum of the particle ${\bf p}$ in $K$ is equal to its momentum ${\bf p}_0$ in $K_0$.
Therefore, the angular momentum of the particle ${\bf M}={\bf r}\times{\bf p}$ in $K$ is equal to its angular momentum ${\bf M}_0={\bf r}\times{\bf p}_0$ in $K_0$.
The energy $E={\bf p}\cdot{\bf v}-L$ for the Lagrangian $L$ is
\begin{equation}
E=\frac{1}{2}m{\bf v}^2-\frac{1}{2}m({\bm\Omega}\times{\bf r})^2+U,
\end{equation}
where the term $-(1/2)m({\bm\Omega}\times{\bf r})^2$ is the centrifugal potential energy.
Putting ${\bf v}={\bf v}_0-{\bm\Omega}\times{\bf r}$ in $E$ gives, using $E_0=(1/2)m{\bf v}^2_0+U$, the relation between the energies $E$ and $E_0$ in the two frames \cite{LL1}:
\begin{equation}
E=\frac{1}{2}m{\bf v}^2_0-m{\bf v}_0\cdot({\bm\Omega}\times{\bf r})+U=E_0-m{\bm\Omega}\cdot({\bf r}\times{\bf v}_0)=E_0-{\bf M}\cdot{\bm\Omega}.
\end{equation}
The relation $E=E_0-{\bf M}\cdot{\bm\Omega}$ is also valid for a system of particles because of the additivity of the Lagrangian.\\

\noindent
{\bf Preferred axis}.\\
The universe seems to be closed \cite{closed}.
In the absence of the rotation of the universe, the number of galaxies that rotate clockwise should be approximately equal to the number of galaxies that rotate counterclockwise.
If the universe rotates with angular velocity ${\bm\Omega}$, the relation $E=E_0-{\bf M}\cdot{\bm\Omega}$ shows that a galaxy with angular momentum ${\bf M}$ has the smallest energy $E$ in the rotating universe if ${\bf M}$ is in the same direction as ${\bm\Omega}$.
Galaxies, as all physical systems, tend to decrease their energies.
In the presence of the rotation of the universe, most galaxies should therefore tend to rotate in a preferred direction, coinciding with the direction of the angular velocity ${\bm\Omega}$ of the universe.
Consequently, the numbers of clockwise- and counterclockwise-spinning galaxies in a rotating universe should be different.

An analysis of the handedness of $\sim 10^4$ spiral galaxies with the redshift $z\sim 0.04$, taken from Sloan Digital Sky Survey, showed a possible asymmetry between clockwise and counterclockwise spins.
The most likely dipole axis, which should coincide with the axis of rotation of the Universe, was found at the right ascension $\alpha=217^\circ$ and declination $\delta=32^\circ$ \cite{gal1}.
An analysis of $\sim 10^5$ spiral galaxies in different regions of the local Universe with $z<0.3$, taken from the same survey, also found a possible spin asymmetry, with the most likely dipole axis at $\alpha=132^\circ, \delta=32^\circ$ \cite{gal2}.
An analysis of $\sim 10^6$ spiral galaxies, taken from the Dark Energy Spectroscopic Instrument (DESI) Legacy Survey, showed a higher number of galaxies spinning counterclockwise in the Northern hemisphere and a higher number of galaxies spinning clockwise in the Southern hemisphere, with the most likely dipole axis at $\alpha=243^\circ, \delta=39^\circ$ \cite{gal3}.
This axis is relatively close to the North Galactic pole, which could be an effect of the motion of the Milky Way.
An analysis of galaxy rotation of $\sim 10^2$ spiral galaxies with $z<2$, taken from JWST Advanced Deep Extragalactic Survey, found a larger asymmetry: the number of galaxies that rotate in the opposite direction relative to the Milky Way galaxy is $\sim50\%$ higher than the number of galaxies that rotate in the same direction relative to the Milky Way \cite{gal4}.
The mean values of the right ascension and declination of the preferred axis are therefore $\alpha=197^\circ\pm 47^\circ, \delta=34^\circ\pm 3^\circ$.

If the universe rotates, the centrifugal force should cause anomalous motion of galaxy clusters in directions perpendicular to the preferred axis and away from it.
An analysis of $\sim 10^2$ early-type galaxies, taken from the Streaming Motions of Abell Clusters project, showed a large-scale bulk flow of galaxy clusters with speed 630 km/s toward $\alpha=128^\circ, \delta=-41^\circ$ \cite{clu1}.
These data are consistent with an analysis of CMB dipole signals taken from the Wilkinson Microwave Anisotropy Probe \cite{clu2}.

The angle between two axes, the first oriented toward a point $P_1$ with $\alpha_1=197^\circ, \delta_1=34^\circ$ and the second oriented toward a point $P_2$ with $\alpha_2=128^\circ, \delta_2=-41^\circ$, can be determined from spherical trigonometry.
The spherical law of cosines for a spherical triangle is
\begin{equation}
\cos c=\cos a\cos b+\sin a\sin b\cos C,
\end{equation}
where $A,B,C$ are vertex angles and $a,b,c$ are the corresponding opposite side angles.
A spherical triangle composed by the great circles connecting the north celestial pole and points $P_1$ and $P_2$ has
\begin{equation}
a=90^\circ-\delta_1,\quad b=90^\circ-\delta_2,\quad C=\alpha_1-\alpha_2.
\end{equation}
Applying the law of cosines to this triangle gives $\cos c=-0.143$, which gives the angle between the two axes: $c=98.2^\circ$.
These axes are nearly perpendicular, as expected.
The alignment of galaxy spins tends to the direction parallel to the preferred axis, whereas the bulk flow of galaxy clusters occurs in directions perpendicular to the preferred axis.
Because the mean right ascension of the preferred axis has a relatively large error, more data are needed to provide definite evidence for this axis.\\

\noindent
{\bf Dark energy as centrifugal force}.\\
Because the universe is closed, the centrifugal force acting in directions perpendicular to the preferred axis manifests itself as a force acting in all directions away from the primordial white hole.
The magnitude of this force becomes $m\Omega^2 r$, where $r$ is the distance of a galaxy from the white hole, instead of $m\Omega^2\rho$.
The corresponding acceleration, $\Omega^2 r$, has the same form as the acceleration $(1/3)\Lambda c^2 r$ resulting from the cosmological constant $\Lambda$.
In an empty universe with a cosmological constant, the corresponding Hubble parameter $H=(\Lambda/3)^{1/2}c$ would thus be equal to $\Omega$.
The cosmological constant generated by the rotation of the universe would not be constant: $\Lambda=3\Omega^2/c^2$.
The Coriolis force, which depends on ${\bm\Omega}$ and ${\bf v}$, introduces additional effects on the motion of galaxies.

Because the angular momentum of the universe is conserved, the angular velocity of the universe decreases as the universe expands.
Consequently, the cosmological constant, which is the simplest explanation of the observed acceleration of the universe expansion (dark energy), should also decrease (evolving dark energy \cite{Avi}).
This result is consistent with recent observations by Dark Energy Survey, showing that dark energy seems to become weaker as the universe expands \cite{DES}.

The origin of dark energy may therefore be explained by the universe formed in a rotating black hole.
This effective cosmological constant, which decreases in time, is not related to quantum field theory or other fundamental physics.
Also, dark energy does not need obscure modifications of general relativity, such as $f(R,\dots)$ gravity, which are merely mathematical extensions of the Einstein equations of general relativity without a physical justification, as discussed in \cite{Sabine}.
The effective cosmological constant may be simply related to the angular velocity of the black hole that created the universe.
Accordingly, different black holes form different universes, which have different dark energies.

I am grateful to Francisco Guedes and my Parents, Bo\.{z}enna Pop{\l}awska and Janusz Pop{\l}awski, for inspiring this work.

\end{document}